\documentclass[aps,pre,showpacs,reprint,superscriptaddress]{revtex4-1}

\usepackage{amsmath,amsfonts,amssymb}
\usepackage[english]{babel}
\usepackage{graphicx}
\usepackage{color}

\begin{document}

\title{Classical dynamics of a two-species condensate driven by a quantum field}

\author{B. M. Rodr\'{\i}guez-Lara}
\affiliation{Institute of Photonics Technologies, National Tsing-Hua University, Hsinchu 300, Taiwan}
\affiliation{Centre for Quantum Technologies, National University of Singapore, 2 Science Drive 3, Singapore 117542}
\author{Ray-Kuang Lee}
\affiliation{Institute of Photonics Technologies, National Tsing-Hua University, Hsinchu 300, Taiwan}
\affiliation{Department of Physics, National Tsing-Hua University, Hsinchu 300, Taiwan}

\begin{abstract}
We present a stability analysis of an interacting two-species Bose-Einstein condensate driven by a quantized field in the semi-classical limit. Transitions from Rabi to Josephson dynamics are identified depending on both the inter-atomic interaction to field-condensate coupling ratio and the ratio between the total excitation number and the condensate size.
The quantized field is found to produce asymmetric dynamics for symmetric initial conditions for both Rabi and Josephson oscillations.
\end{abstract}


\maketitle

\section{Introduction}
Recently, a classical bifurcation at the transition from Rabi to Josephson dynamics has been observed experimentally in a rubidium spinor-Bose-Einstein condensate (BEC) driven by an electromagnetic field through a two-photon coupling process~\cite{Zibold2010p204101}.
A simplified model to access the collective dynamics of a similar full quantum system is that consisting of $N_{q}$ interacting qubits driven by a quantum field through a one-photon process and given by the Hamiltonian,
\begin{eqnarray} \label{eq:ExtDicke}
\hat{H} = \delta \hat{J}_{z} + \eta N_{q}^{-1} \hat{J}_{z}^{2} + \lambda N_{q}^{-1/2} \left(\hat{a}  + \hat{a}^{\dagger} \right) \hat{J}_{x},
\end{eqnarray}
where the detuning $\delta$ is the difference between the BEC ground state hyperfine transition frequency $\omega$ and the quantized field frequency $\omega_{f}$, $\delta = \omega - \omega_f$. The orbital angular momentum representation, with the population difference and coherences given by $\hat{J}_{z}$, $\hat{J}_{x}$ and $\hat{J}_{y}$, in that order, has been chosen to describe the ensemble of dipole-dipole interacting qubits, while the field is described by the creation (annihilation) operator, $\hat{a}$ ($\hat{a}^{\dagger}$). The field-ensemble coupling and the intra-ensemble coupling are given by the constants $\lambda$ and $\eta$, respectively.

A standing conjecture states that entanglement in nonlinear bipartite systems can be associated with a fixed-point bifurcation in the classical dynamics~\cite{Hines2005}, thus providing a link between classical and quantum regimes. 
Entanglement is a fundamental quantum phenomenon~\cite{Schrodinger1935p807} and resource~\cite{Nielsen2000}. 
Interactions are essential to generate quantum correlations; \textit{e.g.}, the Lipkin-Meshkov-Glick (LMG) model, originally proposed in nuclear physics~\cite{Lipkin1965p188}, produces maximal pairwise entanglement of qubits at a quantum phase transition of its ground state and may describe the Josephson effect in a two-mode BEC~\cite{Vidal2004p062304}.

The Hamiltonian in Eq.\eqref{eq:ExtDicke}, hereby called LMG-Dicke (LMGD), can be obtained from the Gross-Pitaevskii equation describing a two-species BEC interacting with a quantum field by following a derivation similar to that found in Ref.~\cite{Chen2007p40004}. 
Experimental realizations providing an assorted range of tunable parameters for the LMGD model may include a two-hyperfine-structure-defined-modes BEC coupled to a quantum cavity field mode through a one microwave photon process; \textit{e.g.} trapped hyperfine ground states of a Sodium BEC inside a microwave cavity~\cite{Gorlitz2003}. 
Arrays of interacting superconducting qubits coupled to the quantum field mode of a coplanar waveguide resonator may be considered with the limitation that ensemble sizes are small~\cite{Tsomokos2010}. 

The ground state phase transition of the LMGD model has been studied in the thermodynamic limit, $N_{q} \rightarrow \infty$, within the rotating wave approximation (RWA)~\cite{Chen2007p40004} and in the quantum regime, using coherent states for both the field and ensemble, without the RWA \cite{Chen2010p053841}.
These results show the existence of a finite size first order quantum phase transition and a second order super-radiant phase transition. In the quantum regime, we have shown~\cite{RodriguezLara2010} that maximal shared bipartite concurrence of the ensemble may be obtained for weak coupling, as the LMGD Hamiltonian in the limit $\lambda \rightarrow 0$ becomes the LMG model.

Here, we present a steady state analysis of the equations of motion for the system in the large ensemble size limit to explore its collective dynamics.
First, the relation between the quantum and classical field drives are discussed comparing the Rabi and Josephson oscillations that appear in both cases.
Then, we find that the symmetry of Josephson dynamics is broken by the quantum field. 
Finally, an actual pitchfork bifurcation point is found in the regime where the intra-ensemble interaction is larger than the field-ensemble coupling, $\eta \gg \lambda$; maximal entanglement is found in this regime for the quantum analysis~\cite{RodriguezLara2010}.
Our results may provide a deeper understanding about the collective dynamics of interacting qubits and another example in favor of the aforementioned conjecture relating classical and quantum regimes for nonlinear systems.

\section{Model}
Starting from the LMGD Hamiltonian defined in Eq.\eqref{eq:ExtDicke}, by considering a large ensemble, $N_{q} \gg 1$, it is possible to approximate the expectation values using the thermodynamic limit where the system is considered to be in a separable state composed by a coherent field, $\vert \sqrt{n} e^{\imath \phi} \rangle$, and coherent atomic state, $\vert \theta, \varphi\rangle$; \textit{i.e.}, $\langle \hat{a}^{\dagger} \hat{a} \rangle = n$, $\langle \hat{a}_{\pm} \rangle = \sqrt{n} e^{\mp \imath \phi}$, $\langle \hat{J}_{z} \rangle \approx (N_{q}/ 2) \cos \theta $, $\langle \hat{J}_{\pm} \rangle \approx (N_{q}/ 2) \sin \theta e^{\pm \imath \varphi}$, and $\langle \hat{N} \rangle = N \approx n - (N_{q} / 2) \cos \theta$.
By defining a fractional population difference, $z = \cos \theta$, an excitation ratio parameter, $k = 2 N / N_{q}$,  and a total phase variable, $\Phi = \phi + \varphi$, the effective Hamiltonian in this mean-field approximation, under the RWA in units of $\hbar  N_{q} \lambda /2$, is given by the expression,
 \begin{eqnarray} \label{eq:Heff-meanfield}
H &=&  \left( \Delta  + \frac{\Lambda z}{2} \right) z  + \left[2 (1 - z^2 )\left(k - z \right) \right]^{1/2} \cos \Phi, 
\end{eqnarray}
where the dimensionless couplings ratio $\Lambda = \eta / \lambda$, detunning $\Delta = \delta / \lambda$, and the shorthand notation for the mean value of an operator $O \equiv \langle \hat{O}\rangle$ has been used.

The mean field Hamiltonian in Eq.~\eqref{eq:Heff-meanfield} describes a non-rigid non-linear pendulum and is equivalent to the case of a BEC in an asymmetric double well via a phase $\pi$-shift and a restriction given by $\left(k - z \right) = 1/2$, \textit{cf.} Eq.(5) in \cite{Smerzi1997p4950}. Under these restrictions, plus symmetry of the double well, $\Delta=0$, the model describes a Bose-Josephson junction supporting coherent oscillations, \textit{i.e.}, Rabi dynamics for macroscopic self-trapping modes and  Josephson dynamics below and above a critical tunneling strength given by $\Lambda_{c} = 1$ \cite{Smerzi1997p4950}. Due to the phase shift with respect to the Bose-Josephson junction, the so-called plasma and $\pi$ oscillations \cite{Raghavan1999p620}, already observed experimentally for bosonic BECs \cite{Zibold2010p204101, Albiez2005p010402}, will exchange places in the studied system; \textit{i.e.} plasma and $\pi$ oscillations will be located at $\Phi = \pi$ and $\Phi = 0$, in that order.

Notice that under resonant quantum driving, $\delta=\omega-\omega_f=0$ leading to $\Delta=0$, all deviations from the symmetric classical driven system are due to the quantized field; the term $(k-z)^{1/2}$ in Eq.\eqref{eq:Heff-meanfield} depends on the parameter $k$, \textit{i.e.} the ratio between the conserved total number of field and atomic excitations in the system and the size of the condensate.

\section{Fixed points}
\begin{figure}
\includegraphics[width= 2.8in]{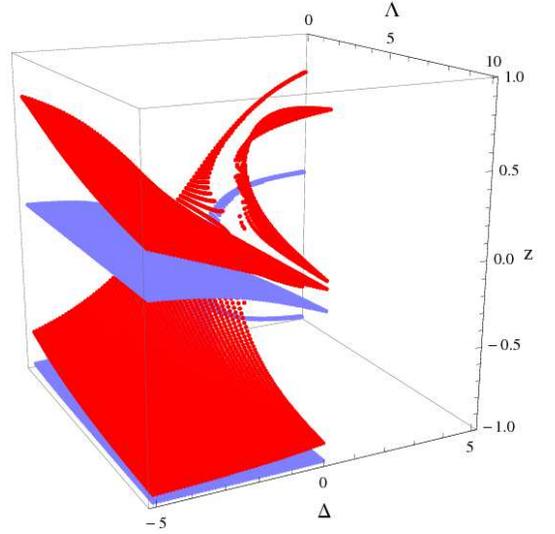}
\caption{(Color Online) Fixed points, $F_{\Phi}$ with $\Phi = 0,~ \pi$, defined by the parameters set given by the dimensionless detunning, couplings ratio, fractional population difference and excitation ratio, $\left\{ \Delta, \Lambda, z, k \right\}$ in that order, for excitation ratio parameters  $k=0.1$ (light blue) and $k=10$ (dark red).   } \label{fig:Figure1}
\end{figure}
\begin{figure}
\includegraphics[width= 3.25in]{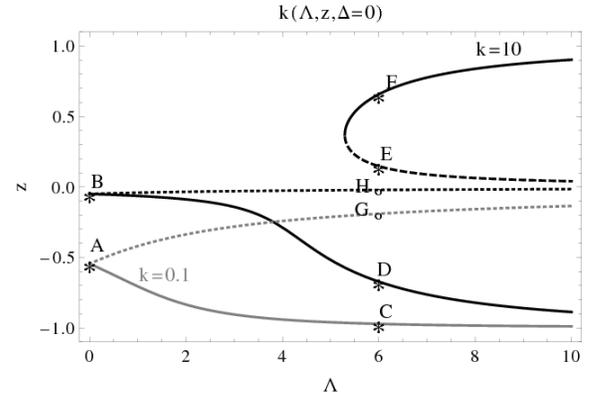}
\caption{Fractional population difference, $z$, as a function of the coupling ratio, $\Lambda$, for two excitation ratios, $k = 0.1$ (grey) and $k=10$ (black) for fixed points $F_{0}$ (solid) and $F_{\pi}$ (dotted). Rabi oscillations occur around the fixed points marked from A to C, Josephson oscillations around fixed points D and F, finally, plasma oscillations around those marked G and H.  The dashed branch for the latter excitation ratio, $k=10$, shows the fixed points acting as separatrix of the two localized oscillations in the Josephson regime, in particular the fixed point E acts as a separatrix of oscillations around fixed points D and F.} \label{fig:Figure2}
\end{figure}

The fixed points of the model, Eq.\eqref{eq:Heff-meanfield}, are found from the mean-field version of the quantum equations of motion, up to a $N_{q}/2$ scale factor, 
\begin{eqnarray} \label{eq:EqMotmf}
\frac{d}{dt} z &=&  \left[2 (1 - z^2)\left( k - z \right) \right]^{1/2} \lambda \sin \Phi, \nonumber \\
\frac{d}{dt} \Phi &=&   \delta +  \kappa z - \frac{\lambda}{\sqrt{2}}\frac{ 1 +  2 k z - 3 z^2}{ \left[ (1 - z^2 )\left(k - z \right) \right]^{1/2}} \cos \Phi . 
\end{eqnarray}
The fixed points coincide with the critical points as $\dot{z} \equiv \partial H / \partial \Phi$ and $\dot{\Phi} \equiv \partial H / \partial z$. Stationary states are found for the phase variable values $\Phi = 0, \pi$ and the excitation parameter value
\begin{eqnarray}\label{eq:ExcParam}
k &=& \frac{3 z^2 - 1}{2 z} + \frac{ (1-z^{2})\vert(\Delta+  \Lambda z) \vert}{4 z^2}\nonumber \\ &&\times \left\{ \vert \Delta +  \Lambda z \vert \pm \left[ \left( \Delta +  \Lambda z \right)^2 - 4  z \right]^{1/2} \right\}.
\end{eqnarray}
Notice that, in order to obtain a real excitation ratio, $k$, the allowed fractional population difference is bounded to the range $z \in [-1, z_{-}] \cup [z_{+},1]$, where $z_{\pm} = [ 2 -   \Delta \Lambda \pm  2 (1 - \Delta \Lambda)^{1/2} ] /  \Lambda^2 $ sets the condition $\Delta \le 1 /  \Lambda$. Requiring population inversion, $J_{z}  = N_{q}/2$, that is $z=1$, leads to $k = 1$,  \textit{i.e.} $N = N_{q}/2$ which means there should be no excitations on the field, $n=0$, without restriction on the couplings ratio $\Lambda$. In a similar way for $J_{z}= - N_{q}/2$: $z=-1$, $k = -1$, $N = - N_{q}/2$, $n=0$. Notice that $z_{\pm}=0$ requires $\Lambda$ or $k \rightarrow \infty$.

Figure \ref{fig:Figure1} shows the fixed points, denoted as $F_{\Phi}$ with $\Phi = 0,~ \pi$, defined by the parameters set $\left\{ \Delta, \Lambda, z, k \right\}$, for excitation parameters $k=0.5$ and $k=10$, \textit{i.e.} the total excitations number is smaller than the number of qubits and \textit{vice-versa}, in that order. Figure \ref{fig:Figure2} shows the plane defined by the resonant case, $\Delta=0$, from Fig. \ref{fig:Figure1}. The different alphabetical markers in Fig.~\ref{fig:Figure2} pinpoint regions with different dynamics. The markers A and B are located in a region where there are only two fixed points per parameter set and these fulfill $z_{F_{0}} = z_{F_{\pi}}$. The next set of markers belongs to a region where two different sets of fixed points are identified; the one for a low excitation ratio where there is one fixed point for plasma and $\pi$ oscillations each, $z_{F_{0}} \ne z_{F_{\pi}}$; the other for a large excitation ratio where there is only one fixed point for $\Phi = \pi$, and three for $\Phi=0$, \textit{i.e.}  a transition from Rabi to Josephson dynamics occurred for $\pi$ oscillations.

\section{Resonant quantum driving examples, $\Delta=0$.}
\begin{figure}
\includegraphics[width= 3.25in]{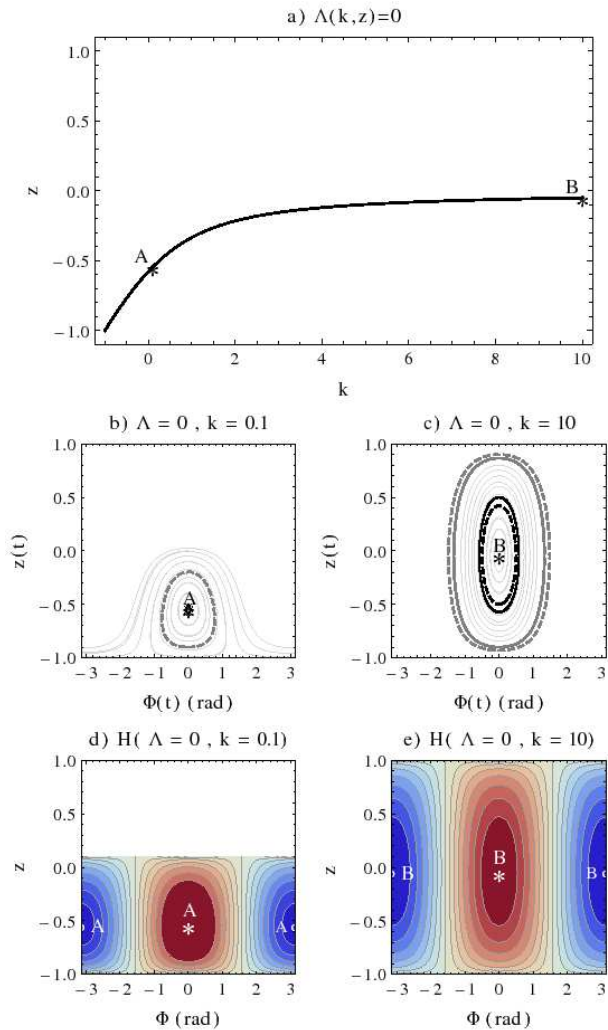}
\caption{(Color online) On resonance, $\Delta=0$, (a) Fractional population difference, $z$, as a function of the excitation ratio, $k$, for a given coupling ratio, $\Lambda=0$. Alphabetical markers show fixed points for excitation ratios $k=0.1$ (A) and $k=10$ (B). (b-c) Trajectories for $\pi$ oscillations around the fixed points A and B for initial conditions including a survey of initial fractional population difference and a initial phase $\Phi(0)=0$; highlighted trajectories sharing the same tone correspond to symmetric initial conditions  $z(0)=0.9$ (dotted gray),  $z(0) = 0.5$ (solid black), $z(0)=-0.5$ (dotted black), $z=-0.9$ (solid gray). (d-e) Mean value for the Hamiltonian showing the critical points defined by A and B. } \label{fig:Figure3}
\end{figure}
\begin{figure}
\includegraphics[width= 3.25in]{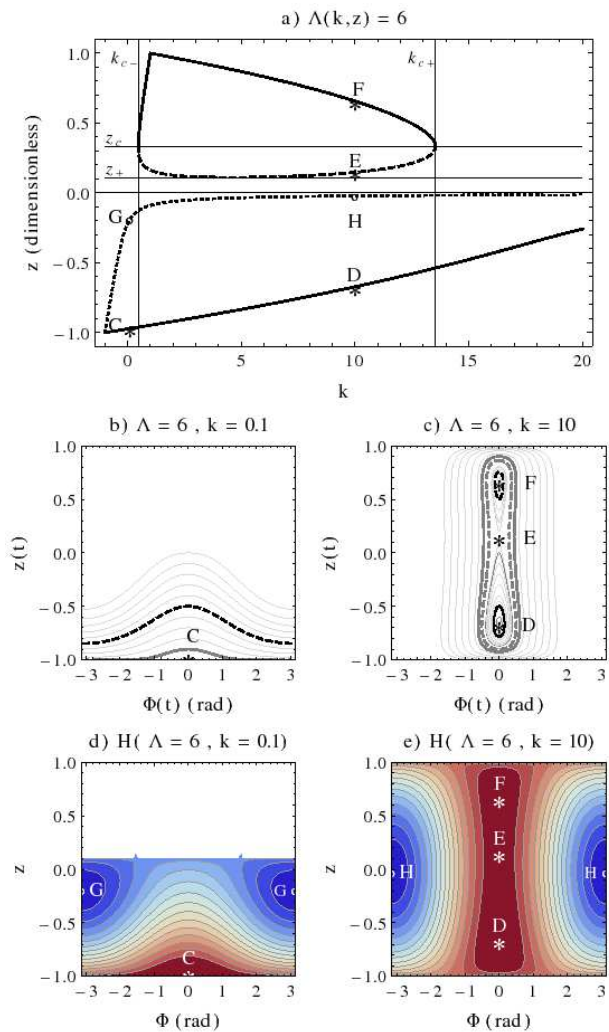}
\caption{(Color online) On resonance, $\Delta=0$, (a) Fractional population difference, $z$, as a function of the excitation ratio, $k$, for a given coupling ratio, $\Lambda=6$. Alphabetical markers show fixed points for excitation ratios $k=0.1$ (C,G) and $k=10$ (D-F,H). (b-c) Trajectories for $\pi$ oscillations around the fixed points C, for low excitation ratio, and D, for large excitation ratio, for initial conditions including a survey of initial fractional population difference and a initial phase $\Phi(0)=0$; highlighted trajectories sharing the same color correspond to symmetric initial conditions  $z(0)=0.9$ (dotted gray),  $z(0) = 0.5$ (solid black), $z(0)=-0.5$ (dotted black ), $z=-0.9$ (solid gray). (d-e) Mean value for the Hamiltonian showing the critical points defined by (C-H).   } \label{fig:Figure4}
\end{figure}

Figure \ref{fig:Figure3} shows an example where Rabi dynamics dominate, \textit{i.e.} $\Lambda < 1$. The coupling ratio $\Lambda=0$ is chosen to explore the behaviour of the system when the inter-qubit coupling is negligible. The addition of the quantized field breaks the symmetry in the trajectories of the equivalent classical field driven case \cite{Lee2009p070401}. Rabi oscillations localize around the fixed point A in the Southern Hemisphere of the Bloch sphere defined by $(\Phi(t), z(t))$; a sample of trajectories with initial conditions on $\Phi(t=0) = 0$, are shown in Fig. \ref{fig:Figure3}(b) where two particular initial conditions, $z=-0.5$ and $z=-0.9$, are highlighted; a few of the trajectories show an unbounded phase. For a larger excitation ratio, fixed point B in Fig. \ref{fig:Figure3}(a), Rabi oscillations for symmetric initial conditions are still asymmetric as shown by the highlighted trajectories in dashed and solid lines portrayed in Fig.\ref{fig:Figure3}(b); it is only in the limit of large excitation ratio, $k \rightarrow \infty$, that symmetric dynamics for symmetric initial conditions are recovered. Notice that unbounded phase modes disappear. Figures \ref{fig:Figure3}(d-e) show the mean value of the mean-field Hamiltonian, Eq.\eqref{eq:Heff-meanfield}, where it is possible to see that $z_{F_{0}} = z_{F_{\pi}}$ for both cases of excitation ratios.

On resonance, $\Delta=0$, for any given value of the coupling ratio, $\Lambda$, the allowed fractional population difference is in the range $ z \in [-1,0) \cup [ 4 / \Lambda^2, 1]$; \textit{i.e.} all of the stationary states are self-trapped states, $z \ne 0$, unless $\Lambda \rightarrow \infty$. For the minimum self-trapped positive fractional population difference, $z_{+} = 4/  \Lambda^2 $, the excitation ratio is given by $k(z_{+}) = ( 16 + \Lambda^4 ) / 8 \Lambda^2$. Furthermore, setting $\partial k / \partial z = 0$ delivers the cubic, with one real root,
\begin{equation}
\Lambda^2 z_{c}^3 - 3 z_{c}^2 -1 = 0,
\end{equation}
that yields the set of critical parameters $\{ k_{c\pm}(\Lambda,z_{c}), z_{c} \}$ delimiting the Josephson dynamics region where there are localized oscillations around two fixed points with positive and negative fractional populations. Figure \ref{fig:Figure4}(a) shows a graph of the fractional population difference, $z$, as a function of the excitation parameter, $k$, for an on-resonance coupling ratio $\Lambda = 6$, where the critical parameters $\{ k_{c}, z_{c} \}$ are shown. For a low excitation parameter, $k=0.1$, coherent oscillations localize in the Southern hemisphere around the fixed point C, Fig.\ref{fig:Figure4}(a); most of these Rabi oscillations are unbounded in phase, Fig.\ref{fig:Figure4}(b). Plasma oscillations localize around the fixed point G, which is the minimum of the mean energy, Eq.\eqref{eq:Heff-meanfield}, Fig.\ref{fig:Figure4}(d). For a larger excitation parameter, $k=10$, an extra branch of fixed points appears and Josephson dynamics showing two self-trapped oscillation modes around fixed points D and F occur, Fig. \ref{fig:Figure4}(a). A  sampler of the trajectories for initial conditions along $\Phi(t=0)=0$, Fig.\ref{fig:Figure4}(c), shows that the fixed point E is the separatrix of the two sets of self-trapped oscillations. Figure \ref{fig:Figure4}(d) shows the mean energy for the system and how the fixed points correspond to local maxima, D and F, minima, H, and saddle, E, points. The localization of the self-trapped modes is asymmetric; symmetric dynamics for symmetric initial conditions can be recovered only in the limit $\Lambda \rightarrow \infty$. 

\begin{figure}
\includegraphics[width= 3.25in]{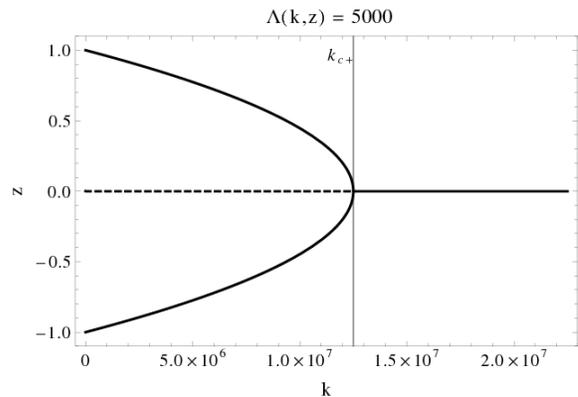}
\caption{On resonance, $\Delta=0$, fractional population difference, $z$ (solid line), as a function of the excitation ratio, $k$, for a very large coupling ratio, $\Lambda=5000$. A bifurcation point appears at the critical excitation ratio $k_{c+} \approx \Lambda^2 /2 = 1.25 \times 10^7$. The separatrix for the Josephson regime is shown as a dashed line.   } \label{fig:Figure5}
\end{figure}

Figure (\ref{fig:Figure5}) shows that for large coupling ratios, $\Lambda \gg 1$, \textit{i.e.} $\eta \gg \lambda$, it is possible to locate a pitchfork bifurcation point, $k_{c+} \approx \Lambda^2/2$. This condition, $\eta \gg \lambda$, relates to the phase space region where maximal shared bipartite  concurrence in the qubit ensemble may be obtained in the quantum treatment of this model \cite{RodriguezLara2010}. The difference comes from the large excitation parameter ratio arising in this semi-classical analysis, $k_{c+} \gg 1$, \textit{i.e.} $N_{q} \ll n$ as $z_{c} \approx 3 / \Lambda^2 \ll 1$.

\section{Discussion}

In the given examples, we have shown how the ratio between the total number of excitations of the system and the condensate size, $k= 2N / N_{q}$, limit the region of phase space accessible to both Rabi and Josephson oscillations.
Only in the limit when the total number of field and atomic excitations in the system is infinitely larger than the size of the condensate, $N \gg N_{q}/2$ leading to $k \rightarrow \infty$, the whole phase space is available for the oscillations. As the value of $k$ diminishes, \textit{i.e.} the number of field and atomic excitations become less, the available region of phase space reduces, leading to more trajectories unbounded in phase for Rabi oscillations.

The excitation ratio parameter, $k$, may be highly tunable as it relates to the population of the atomic species and the number of photons in the quantized driving field, all of them highly controllable in contemporaneous experimental realizations.

More interesting is the asymmetric bifurcation of the nonlinear dynamics, with the displacement of the separatrix from the typical value of $z=0$, for values close and above the critical coupling ratio $\Lambda_{c} = 1$. 
A symmetric pitchfork bifurcation is recovered for an infinitely large coupling ratio, $\Lambda = \eta/\lambda \rightarrow \infty$; \textit{i.e.}, the LMGD Hamiltonian, Eq.\eqref{eq:ExtDicke},  becomes simply a LMG model, $\lambda \rightarrow 0$. 
Hence the breaking of this particular nonlinear symmetry comes as a result of the quantized field driving. 
The behavior of the asymmetric bifurcation may be proved by varying the inter-atomic interaction, $\eta$, with respect to the atomic-field coupling, $\lambda$, with the condition that $\eta > \lambda$ and the extra restrictions provided above to see Josephson dynamics. 

Notice that both the inter-atomic interaction and the atomic-field coupling are highly tunable parameters in current experimental realizations. Furthermore, in this particular scheme the weak coupling between a hyperfine transition and a quantized microwave field within the RWA is desirable to explore large values of $\Lambda$.

\section{Conclusion}
In summary, we have presented a stability analysis of the classical dynamics of a large ensemble of interacting qubits driven by a quantized field. There exists a transition from Rabi to Josephson oscillations as in the classical field driven system. The quantized field produces localized asymmetric dynamics for symmetric initial conditions while the classical field produces symmetric dynamics for symmetric initial conditions. For low coupling ratios, depending on the excitation ratio, Rabi oscillations localize in the Southern Hemisphere of the corresponding Bloch sphere for low excitation ratios, and smoothly transit to cover both Hemispheres for large excitation ratios; completely identical coherent oscillations for symmetric initial conditions are only recovered for infinitely large excitation ratios. For large coupling ratios, there exists a critical excitation ratio that defines two different sets of dynamics; below this critical excitation ratio, localized coherent oscillations, most of them showing an unbounded phase, appear in the Southern hemisphere of the Bloch sphere; above the critical excitation ratio, two localized, or self-trapped, oscillation modes characteristic of Josephson dynamics appear with asymmetric dynamics for symmetric initial conditions and a separatrix with a non-null fractional population difference. A pitchfork bifurcation, somehow similar to the classical field driven case, is found for large coupling and excitation ratios, in these cases the separatrix corresponds to ensemble states with negligible fractional population which are closer to their classical field driven equivalent. 

\begin{acknowledgments}
The authors are indebted to Soi-Chan Lei and Chaohong Lee for useful discussions and to the anonymous referees for their clarifying comments.
\end{acknowledgments}

\end{document}